\documentclass[prb,aps,twocolumn,showpacs,nobibnotes,epsf]{revtex4}

\usepackage{graphicx}
\usepackage{dcolumn}
\usepackage{bm}
\usepackage{SIunits}

\begin{document}
\title{Transport and magnetic properties of La-doped CaFe$_2$As$_2$}
\author{J. J. Ying, J. C. Liang, X. G. Luo, X. F. Wang, Y. J. Yan, M. Zhang, A. F. Wang, Z. J. Xiang, G. J. Ye,
P. Cheng and X. H. Chen} \altaffiliation{Corresponding author}
\email{chenxh@ustc.edu.cn} \affiliation{Hefei National Laboratory
for Physical Science at Microscale and Department of Physics,
University of Science and Technology of China, Hefei, Anhui 230026,
People's Republic of China\\}

\begin{abstract}
We measured the transport properties and susceptibility of single
crystals Ca$_{1-x}$La$_x$Fe$_2$As$_2$(x=0, 0.05, 0.1, 0.15, 0.19 and
0.25). Large in-plane resistivity anisotropy similar to that in
Co-doped 122 iron-pnictides is observed although no transition
metals were introduced in the FeAs-plane. The in-plane resistivity
anisotropy gradually increases with La doping below T$_{SDW}$, being
different from the hole-doped 122 superconductors. The
susceptibilities of the samples show that La doping leads to
suppression of SDW and induces a Curie-Weiss-like behavior at low
temperature, which is much stronger than the other 122 iron-based
superconductors.
\end{abstract}


\vskip 300 pt

\maketitle Iron-based superconductors have attracted great
attentions these years\cite{Kamihara,chenxh, ren, rotter}. The
parent compound undergoes structure and spin density wave (SDW)
transitions. With chemical doping or high pressure, both structure
and SDW transition can be suppressed and superconductivity emerges.
$A$Fe$_2$As$_2$ ($A$=Ca, Sr, Ba, Eu, so called "122") with the
$ThCr_2Si_2$-type structure were widely investigated because it is
easy to grow large size of high quality single crystals. However,
the highest Tc of 122 superconductors do not surpass 40 K in the
earlier studies. Recently, superconductivity up to 49 K was
discovered in rare earth doped CaFe$_2$As$_2$\cite{Saha, Gao, Lv,
Qi}. The FeAs plane of Ca$_{1-x}$$R$$_x$Fe$_2$As$_2$($R$=rare earth
elements) is not affected by substitution of trivalent $R^{3+}$ ions
on divalent Ca$^{2+}$. While for the other two famous electron-doped
122 superconductors BaFe$_{2-x}$Co$_x$As$_2$ and
K$_x$Fe$_{2-y}$Se$_2$, some of the Fe ions in the FeAs/FeSe layers
are either substituted by Co or missed\cite{Sefat, xlchen, Ying}. It
is an ideal candidate for us to investigate the electron-doped
iron-based superconductors with the perfect FeAs layers. The
superconducting temperature in Ca$_{1-x}$$R$$_x$Fe$_2$As$_2$ is much
higher compared to the other 122 iron pnictides and it is very
necessary for us to detailedly investigate the physical properties
of Ca$_{1-x}$$R$$_x$Fe$_2$As$_2$.

\begin{figure}[t]
\centering
\includegraphics[width=0.5 \textwidth]{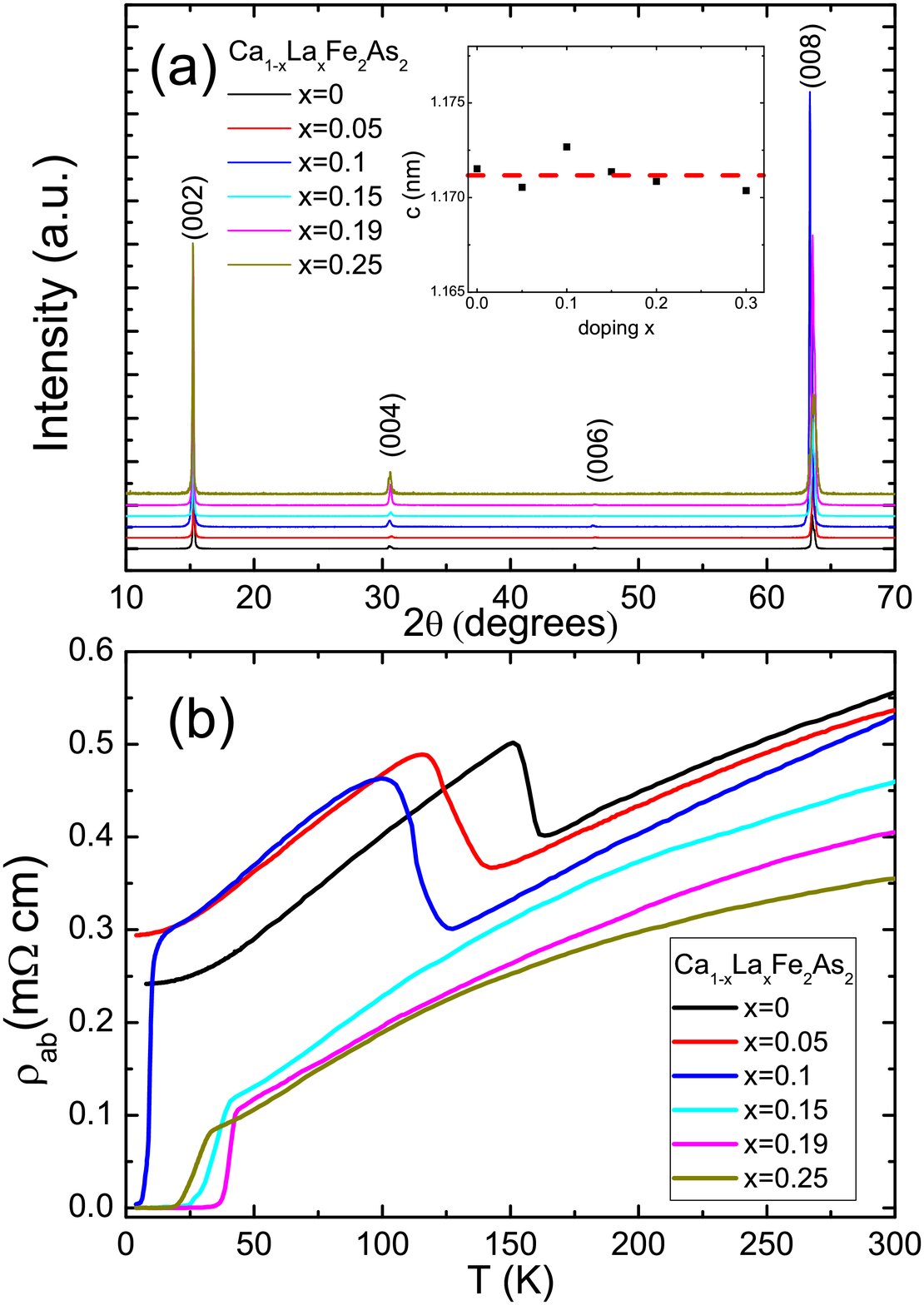}
\caption{(Color online)(a): The single crystal x-ray diffraction
pattern of Ca$_{1-x}$La$_x$Fe$_2$As$_2$, Only (00$l$) diffraction
peaks show up, indicating that the c axis is perpendicular to the
plane of the plate. The inset shows that the c parameter does not
changes with doping. (b): Temperature dependence of the in-plane
resistivity.} \label{fig1}
\end{figure}

\begin{figure}[t]
\centering
\includegraphics[width=0.5 \textwidth]{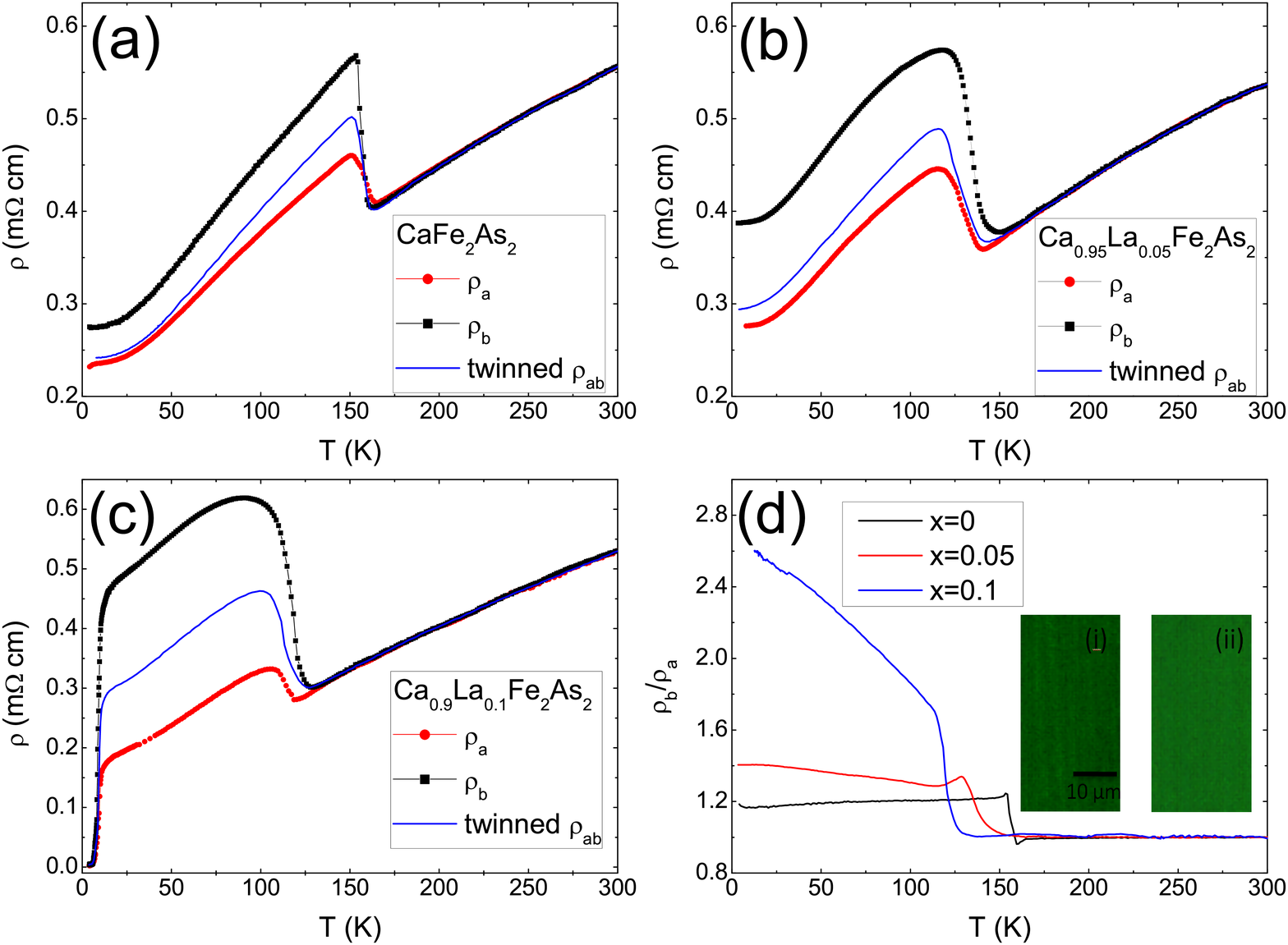}
\caption{(Color online) Temperature dependence of in-plane
resistivity with the electric current flow along $a$ direction and
$b$ direction respectively for (a): parent compound CaFe$_2$As$_2$,
(b): Ca$_{0.95}$La$_{0.05}$Fe$_2$As$_2$, (c):
Ca$_{0.9}$La$_{0.1}$Fe$_2$As$_2$. The twinned in-plane resistivity
was also shown for comparison (blue line). (d): Temperature
dependence of in-plane resistivity anisotropy $\rho_b$/$\rho_a$. The
insets of (f) are the Polarized-light images of the surface for
twinned(i) and detwinned(ii) CaFe$_2$As$_2$ at the temperature of 78
K.} \label{fig2}
\end{figure}

\begin{figure}[t]
\centering
\includegraphics[width=0.5 \textwidth]{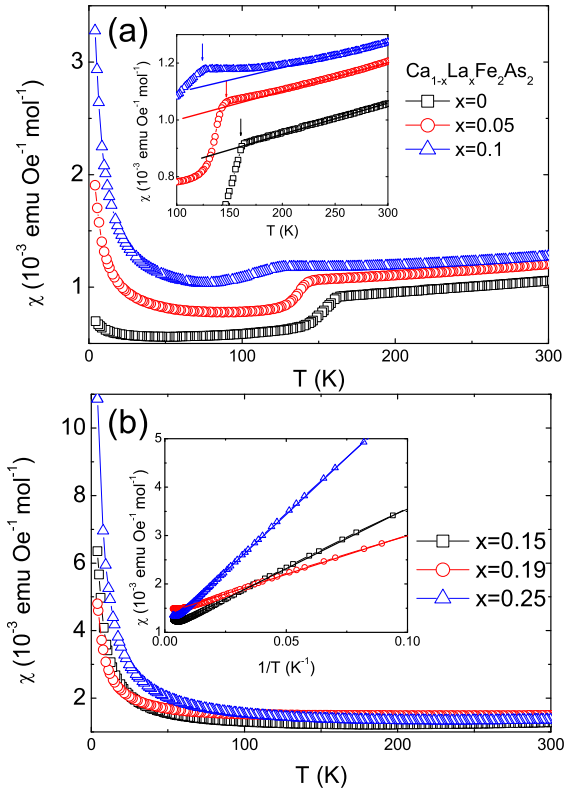}
\caption{(Color online) (a): Temperature dependence of magnetic
susceptibility for x=0, 0.05 and 0.1. The inset is the enlarged area
around $T_{SDW}$. The arrows indicate the $T_{SDW}$. (b):
Temperature dependence of magnetic susceptibility for x=0.15, 0.19
and 0.25. The inset shows that the magnetic susceptibility is linear
with 1/T at low temperature, indicating a nice Curiew-Weiss behavior
in the low temperature.} \label{fig4}
\end{figure}

\begin{figure}[t]
\centering
\includegraphics[width=0.5 \textwidth]{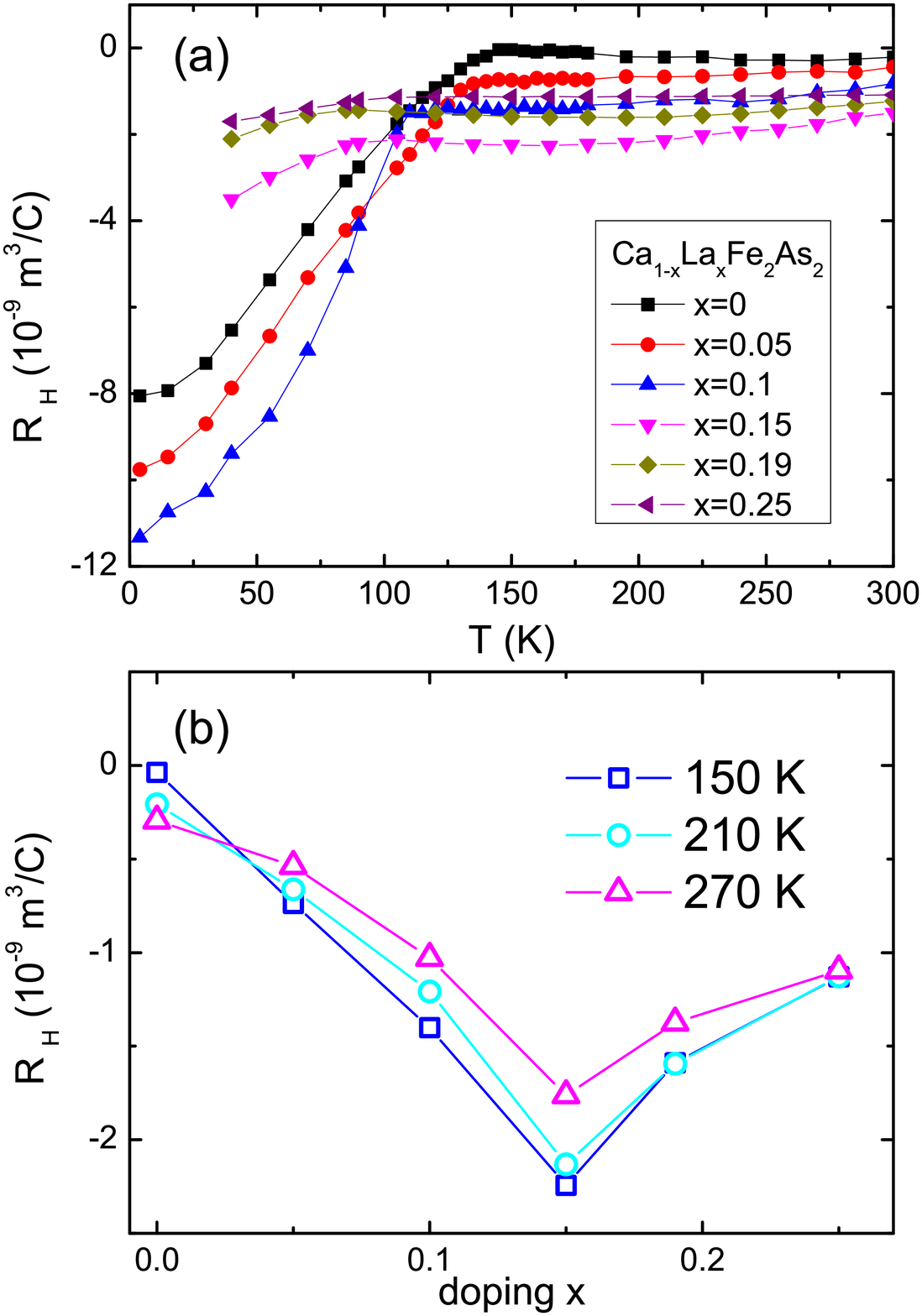}
\caption{(Color online) (a): Temperature dependence of R$_H$ for
Ca$_{1-x}$La$_x$Fe$_2$As$_2$. (b): Doping dependence of R$_H$ at
certain temperature in the normal state.} \label{fig4}
\end{figure}

Recent works showed large in-plane resistivity anisotropy below
$T_S$ or $T_N$ in the parent and electron-doped 122 system although
the distortion of the orthorhombic structure is less than 1
\%\cite{Chu, Tanatar, Ying2, fisher} in the SDW state. However, for
the hole-doped Ba$_{1-x}$K$_x$Fe$_2$As$_2$, in-plane resistivity
anisotropy is nearly absent\cite{Ying2}. All the previous works were
focusing on the Co, Ni and Cu substitution on the Fe site for
electron-doped samples\cite{Kuo} and it is supposed that the large
in-plane anisotropy might come from the transition metal
substitution in the FeAs plane. It is very meaningful for us to
study the in-plane resistivity anisotropy for the electron-doped
Ca$_{1-x}$$R$$_x$Fe$_2$As$_2$ with the perfect Fe square lattice.
The large in-plane anisotropy observed in
Ca$_{1-x}$$R$$_x$Fe$_2$As$_2$ in this paper indicates that the large
in-plane resistivity anisotropy is related to the electron-doping
rather than transition metal doped in the FeAs plane.

In this paper, we systematically investigated the transport and
magnetic properties of La-doped CaFe$_2$As$_2$. Superconductivity up
to 43 K was observed in the Ca$_{0.81}$La$_{0.19}$Fe$_2$As$_2$
similar to the previous results\cite{Gao}. SDW transition can be
suppressed through La doping and superconductivity coexists with
antiferromagnetic with the La doping level between 0.05 and 0.15.
The in-plane resistivity anisotropy gradually increases with La
doping below T$_{SDW}$, being different from the hole-doped 122
superconductors.  A strong Curie-Weiss-like behavior at low
temperature is induced by La-doping.

High quality single crystals with nominal composition
Ca$_{1-x}$La$_x$Fe$_2$As$_2$(x=0, 0.05, 0.1, 0.15, 0.2 and 0.3) were
grown by conventional solid-state reaction using FeAs as
self-flux\cite{Gao, Lv, Qi}. The FeAs precursor was first
synthesized from stoichiometric amounts of Fe and As inside the
silica tube at 800 $\celsius$ for 24 h.  Appropriate amounts of the
starting materials of FeAs, Ca and La were placed in an alumina
crucible, and sealed in an quartz tube. The mixture was heated to
1180 $\celsius$ in 6 hours and then kept at this temperature for 10
hours, and later slowly cooled down to 950 $\celsius$ at a rate of 3
$\celsius$/ hour. After that, the temperature was cooled down to
room temperature by shutting down the furnace. The shining platelike
Ca$_{1-x}$La$_x$Fe$_2$As$_2$ crystals were mechanically cleaved from
the flux and obtained for measurements. The actual composition of
the single crystals were characterized by the Energy-dispersive
X-ray spectroscopy (EDX). The actually doping levels are almost the
same with the nominal values for x smaller than 0.2. While for the
nominal composition x = 0.2 and 0.3, the actual values of x are 0.19
and 0.25 which are smaller than the nominal values. Resistivity was
measured using the Quantum Design PPMS-9 and Magnetic susceptibility
was measured using the Quantum Design SQUID-MPMS. In-plane
resistivity anisotropy was measured using the same method with the
previous work\cite{Chu,Ying2}. Crystals were cut parallel to the
orthorhombic a and b axes so that the orthorhombic a(b) direction is
perpendicular (parallel) to the applied pressure direction. $\rho_a$
(current parallel to a) and $\rho_b$ (current parallel to b) were
measured on the same sample using standard 4-point configuration.

Single crystals of Ca$_{1-x}$La$_x$Fe$_2$As$_2$ were characterized
by x-ray diffractions (XRD) using Cu $K_\alpha$ radiations. As shown
in the Fig.1(a). Only (00$l$) diffraction peaks were observed,
suggesting that the crystallographic c axis is perpendicular to the
plane of the single crystal. The inset of Fig.1(a) shows the c-axis
parameters with different doping level, we can see that the lattice
parameters of c-axis almost do not change with La doping. The
lattice constant of c-axis was around 11.72 {\AA} for all the
samples. It is slightly different from the polycrystalline samples
reported previously for which the c-axis slightly decrease with rare
earth elements doping\cite{Saha}. Fig.1(b) shows the temperature
dependence of the resistivity with the electric current flowing in
the ab plane for Ca$_{1-x}$La$_x$Fe$_2$As$_2$ single crystals. The
upturn of resistivity for parent compound was ascribed to the SDW or
structure transition. With La doping, the anomaly was gradually
suppressed and superconductivity gradually emerged. For the optimal
doped sample x = 0.19, the onset of superconductivity was up to 43 K
which is almost the same with the earlier results\cite{Gao}.
However, the superconductivity transition in this system is much
broader than that in other 122 iron-pnictide superconductors.

Fig.2(a) shows the temperature dependence of in-plane resistivity
with the current flowing parallel to the orthorhombic $b$ direction
(black) and orthorhombic $a$ direction (red) of the detwinned
CaFe$_2$As$_2$ sample. The insets of Fig.2(d) are the
Polarized-light images of the surface for twinned(i) and
detwinned(ii) CaFe$_2$As$_2$ at the temperature of 78 K. The samples
are almost fully detwinned as we can no longer see the twin domains.
Obvious anisotropy was observed which is similar to the earlier
result\cite{Tanatar}, although different methods were used to detwin
the samples. Compounds with a small amount of La-doping show much
larger in-plane anisotropy as displayed in Fig.2(b) and (c) for
x=0.05 and x=0.1, respectively. The different behavior of $\rho_a$
and $\rho_b$ observed here is very similar to the ones in other
parent or electron-underdoped iron-pnictides\cite{Tanatar, Chu}. We
characterized the degree of in-plane resistivity anisotropy by the
ratio $\rho_b$/$\rho_a$. Fig.2 (d) shows the temperature dependence
of $\rho_b$/$\rho_a$ for different doping level samples. For all the
samples, in-plane resistivity anisotropy increases very quickly
below T$_{SDW}$. With increasing La doping level, the in-plane
resistivity anisotropy gradually increases in the SDW region. It is
very similar with the other underdoped electron-doping 122 systems,
but quite different from the hole-doped
Ba$_{1-x}$K$_x$Fe$_2$As$_2$\cite{Ying2}. Although the doping
position is away from the FeAs plane, large in-plane anisotropy
still exists and increases with doping content. The magnitude of
in-plane resistivity anisotropy is almost the same as the other
underdoped electron-doping 122 samples. These results suggest that
the in-plane anisotropy is closely related to the carrier-type
rather than the different doping positions.

The temperature dependence of magnetic susceptibilities for
Ca$_{1-x}$La$_x$Fe$_2$As$_2$ (x=0, 0.05 and 0.1) with magnetic field
of H = 1T applied along the ab-plane are shown in Fig.3 (a). The
susceptibility of parent compound CaFe$_2$As$_2$ shows T-linear
behavior and gradually decreases with decreasing temperature above
T$_{SDW}$. The susceptibility suddenly drops at T$_{SDW}$ and shows
very weak Curie-Weiss-like behavior at low temperature. This
behavior is the same with the previous result\cite{Wu}. Small amount
doping of La obviously enhances the Curie-Weiss-like behavior at low
temperature. With increasing the doping content, the anomaly due to
SDW order is gradually suppressed and T-linear behavior of
susceptibility is gradually broken due to the Curie-Weiss-like
behavior at low temperature. It strongly contrast to Co-doped 122
samples in which the T-linear behavior of susceptibility survives
with Co doping\cite{Wang}. With further increasing the doping
content, the susceptibilities showed strong Curie-Weiss-like
behavior below about 200 K as shown in Fig.3 (b). Such strong
Curie-Weiss-like behavior observed in Ca$_{1-x}$La$_x$Fe$_2$As$_2$
is strongly different to the other 122 iron-pnictide
superconductors.

Fig.4(a) shows the temperature dependence of Hall coefficient R$_H$
of Ca$_{1-x}$La$_x$Fe$_2$As$_2$. R$_H$ of all the samples show the
negative values, indicating that the dominated carrier is electron.
The sharp increase of the absolute value of R$_H$ below $T_{SDW}$
indicates the sudden drop of carrier density in the SDW state, which
is the common feature in iron pnictide superconductors\cite{Fang}.
It is also instructive to analyze the Hall Coefficient at certain
temperature for different doping level samples at high temperature.
It is found that the dependence of doping is nonmonotonic as shown
in Fig.4(b). For the samples x$<$0.15, the absolute value of R$_H$
increases with increasing the doping content. With further doping,
it gradually decreases. This behavior is quite similar to the
BaFe$_{2-x}$Co$_x$As$_2$ system. Such behavior is caused by the
multiband effect and different mobility of electron or hole
carriers\cite{Fang}.

Large in-plane anisotropy is observed in the underdoped region of
Ca$_{1-x}$La$_x$Fe$_2$As$_2$ although no extra elements were
introduced into the FeAs-plane. It strongly contrasts to the
hole-doped Ba$_{1-x}$K$_x$Fe$_2$As$_2$ system, but is similar to the
transition metal substitution samples\cite{Ying2, Chu}. Our results
show that substitution away from the FeAs plane could also lead to
large in-plane anisotropy, which indicate that the in-plane
anisotropy is strongly dependent on doping carriers rather than the
doping site. The in-plane anisotropy was suggested closely related
to the the orbital degree of freedom\cite{Chen, Weicheng} or/and
spin fluctuations\cite{Fernandes}. Our previous work demonstrates
that the microscopic orbital involvement in magnetically ordered
state must be fundamentally different between the hole and electron
underdoped iron pnictides\cite{Ying2}. Large in-plane anisotropy
observed in rare earth doped materials also prove this. ARPES
experiments need to do to compare the orbital characters of the hole
and electron doped samples. The magnetic susceptibilities of
Ca$_{1-x}$La$_x$Fe$_2$As$_2$ show much stronger Curie-Weiss-like
behavior compared to the other 122 iron pnictide superconductors.
Although superconductivity above 40K was observed in this system
which is much higher than the other 122 superconductors,
superconducting transition is broad. The origins of the strong
Curie-Weiss-like behavior and broad superconducting transition in
this system are still unknown.

In conclusion, we systematically measured the transport properties
and susceptibilities of Ca$_{1-x}$La$_x$Fe$_2$As$_2$(x=0, 0.05, 0.1,
0.15, 0.19 and 0.25). Large in-plane anisotropy is observed which is
similar to Co-doped 122 iron-pnictides, but strongly contrasts to
the hole-doped Ba$_{1-x}$K$_x$Fe$_2$As$_2$. Strong Curie-Weiss-like
behavior of susceptibility was observed at low temperature by
increasing the La doping.

{\bf ACKNOWLEDGEMENT} This work is supported by the National Basic
Research Program of China (973 Program, Grant No. 2012CB922002 and
No. 2011CB00101), National Natural Science Foundation of China
(Grant No. 11190020 and No. 51021091), the Ministry of Science and
Technology of China, and Chinese Academy of Sciences.
\\

\end{document}